\documentclass[aps,prb,showpacs,reprint, superscriptaddress]{revtex4-1}

\usepackage{graphicx}
\usepackage{color}
\usepackage{amsmath}
\usepackage{bbm}
\usepackage{amssymb}

\usepackage[utf8]{inputenc}	
\usepackage[T1]{fontenc}

\input epsf

\begin{document}

\title{Effect of interlayer processes on the superconducting state within $t$-$J$-$U$ model: 
Full Gutzwiller wave-function solution and relation to experiment}
\author{Micha{\l} Zegrodnik}
\email{michal.zegrodnik@agh.edu.pl}
\affiliation{Academic Centre for Materials and Nanotechnology, AGH University of Science and Technology, Al. Mickiewicza 30, 30-059 Krakow,
Poland}
\author{J\'ozef Spa\l ek}
\email{jozef.spalek@uj.edu.pl}
\affiliation{Marian Smoluchowski Institute of Physics, 
Jagiellonian University, ul. \L ojasiewicza 11,
30-348 Krakow, Poland}
\affiliation{Academic Centre for Materials and Nanotechnology, AGH University of Science and Technology, Al. Mickiewicza 30, 30-059 Krakow,
Poland}

\date{13.09.2016}

\begin{abstract}
The Gutzwiller wave function solution of the $t$-$J$-$U$ model is considered for the bilayer high-T$_C$ superconductor by using the so-called diagrammatic expansion method. The focus is on the influence of the interlayer effects on the superconducting state. The chosen pairing symmetry is a mixture of $d_{x^2-y^2}$ symmetry within the layers and the so-called $s^{\pm}$ symmetry for the interlayer contribution. The analyzed interlayer terms reflect the interlayer electron hopping, the interlayer exchange coupling, and the interlayer pair hopping. The obtained results are compared with selected experimental data corresponding to the copper-based compound Bi-2212 with two Cu-O planes in the unit cell. For the sake of comparison, selected results for the case of the bilayer Hubbard model are also provided. This paper complements our recent results obtained for the single-plane high temperature cuprates (cf. Ref. \onlinecite{Spalek2016}).
\end{abstract}


\maketitle
\section{Introduction}

The crystal lattice of high-$T_C$ cuprate superconductors is composed
of blocks of Cu-O planes separated by charge-reservoir layers\cite{Uchida2015}. 
In the case of compounds with two (or more) Cu-O planes within a single block, the so-called bilayer splitting of the Fermi surface appears due to the hybridization of electron states originating from the individual layers. 
Initially, it was predicted that such splitting should vanish in the nodal ($k_x=k_y$) direction. However, the ARPES analysis for Bi-2212\cite{Kordyuk2004} and 
YBCO\cite{Borisenko2006} compounds has shown a nonzero splitting between the hybridized bonding and antibonding bands in that direction. 
Possible explanation of appearance of such a splitting can be the nodal interlayer coupling \cite{Garcia2010}, mixing with the chain bonds \cite{Andersen1995}, or sensitivity of the quasiparticle spectrum in the nodal region to small effects such as the spin-orbit coupling \cite{Harrison2015}. It has also been argued that
the main contribution to the nodal splitting comes from the vertical hopping between the $O$ $2p_{\sigma}$ orbitals\cite{Kordyuk2004}.
At the same time, with the increasing number of layers in a single Cu-O block the superconducting critical temperature is increased
(for the number of layers $n\leqslant 3$)
indicating that the interlayer processes influence the pairing strength. 
Also, the optical Josephson plasma modes measurements\cite{Kleiner1992} show that the interlayer
Josephson coupling strength can be correlated with the value of the critical temperature, $T_C$.

Even though a vast majority of previous theoretical studies regarding the copper based materials
is focused on a single Cu-O plane, the bilayer Hubbard, $t$-$J$, and $t$-$J$-$U$ 
models have also been
considered\cite{Medhi2009,Maier2011,Lanata2009,Mori2006,Nishiguchi2013,Voo2015,Ruger}.
What distinguishes the latter models from the 
single-layer approach 
is, among others, the appearance of the interlayer pairing channels which can contribute to the total
superconducting gap. In this respect, a possible choice of pairing symmetries are 
discussed in Refs. \onlinecite{Medhi2007,Medhi2007_2}. The interalyer pairing leads to deviations from the pure $d$-$wave$ character of the superconducting order parameter what may be important in the context of some of the ARPES
experiments indicating that the gap symmetry is more complex than a simple 
$d$-$wave$ \cite{Zhao2007,Ding1995,Vobornik1999}. 

Another effect analyzed within the bilayer models is the interlayer pair tunneling. It has been shown that the initially proposed interlayer
Josephson coupling arising as a second-order process in the interlayer electron 
hopping\cite{Chakravarty1993} is 
insufficient to enhance significantly superconductivity due to a small value of 
the interlayer hopping $t_z$ with respect to the intralayer value $t$ (such coupling would be $\propto t_z^2/t$). 
The pair tunneling term has also been introduced in a phenomenological manner 
to postulate the form of the in-phase gap function between two layers\cite{Chen2012}. A different approach is to investigate the interlayer pair 
hopping as originating from the matrix elements of the long-range Coulombic interaction\cite{Nishiguchi2013}.

Here we analyze the influence of interalyer processes such as the interlayer electron hopping, the interlayer pair hopping, and the interlayer exchange coupling, on the bulk superconducting state within the $t$-$J$-$U$ model of the bilayer structure. One pf the purposes of such analysis is to see to what extent the universal properties of the single plane systems are preserved. To take into account the strong electronic correlations in the system, we use the \textit{diagrammatic expansion Gutzwiller wave function method} (DE-GWF)\cite{Bunemann2012,Gebhard1990,Kaczmarczyk2013,Kaczmarczyk2014}. The choice of the model has been dictated by a good quantitative agreement between the theoretical results and principal experimental data for the copper-based materials, which has been reported recently for the single layer version of the same model\cite{Spalek2016}. The gap symmetry selected in our analysis is a mixture of $d_{x^2-y^2}$ symmetry within the layer and the so-called $s^{\pm}$ symmetry for the interlayer 
contribution\cite{Maier2011,
Ubbens1994,Lee1995}. The resulting gap function has the form $\Delta(\mathbf{k})=\Delta_d(\mathbf{k})\pm \Delta_{\perp}$ 
with the $+$ ($-$) sign, corresponding to the bonding (antibonding) band and $\Delta_d(\mathbf{k})$ representing the intralayer $d$-$wave$ gap. As reported in Refs. \onlinecite{Lee1995,Ubbens1994}, such a state has been found as stable for the case of the $t$-$J$ model within the slave-boson analysis. In contradistinction to these results, the coexistence of the interlayer and intralayer pairings has not been found within of the variational Monte Carlo (VMC) method for the same model\cite{Maier2011}. Within our analysis of the bilayer $t$-$J$-$U$ model, the admixture of the interlayer $s^{\pm}$ contribution leads to a small lowering of the system energy with the interlayer gap magnitude being one order of magnitude smaller than the intralayer one. As shown earlier\cite{Kaczmarczyk2014} for the case of a single-layer $t$-$J$ model, our results are of the VMC quality.

For the sake of completeness, we present here also the results for the bilayer Hubbard model and analyze the influence of the differentiation between the values of the Hubbard $U$ corresponding to the two layers.

The structure of the paper is as follows. In Section II and III we show explicitly the model Hamiltonian and describe the basic concept behind the DE-GWF method as applied to the bilayer structure. In Section IV we provide the results of our calculations and analyze them. In particular, we compare our results with available experimental data for the dispersion relation, the Fermi velocity, and the bilayer splitting, all in the nodal direction. The summary is provided in the last Section. Note also that the results supplement our previous analysis for a single-layer situation\cite{Spalek2016} that should be regarded as Part I of the series. In that paper the methodology behind selecting the $t$-$J$-$U$ model is also discussed.

\section{Model}
We consider the $t$-$J$-$U$ model for the bilayer high-T$_C$ superconductor represented by the following Hamiltonian
\begin{equation}
\mathcal{\hat{H}}=\mathcal{\hat{H}_{\parallel}}+\mathcal{\hat{H}_{\perp}}\;,
 \label{eq:H_start_general}
\end{equation}
where
\begin{equation}
\mathcal{\hat{H}_{\parallel}}=\sideset{}{'}\sum_{ijl\sigma}t_{ij}\hat{c}^{\dagger}_{il\sigma}\hat{c}_{jl\sigma}
+U\sum_{il} \hat{n}_{il\uparrow}\hat{n}_{il\downarrow}
+\sideset{}{'}\sum_{ ijl}J_{ij}\hat{\mathbf{S}}_{il}\cdot\hat{\mathbf{S}}_{jl}\;,
 \label{eq:H_start_parallel}
\end{equation}
represents the intralayer part and
\begin{equation}
\begin{split}
\mathcal{\hat{H}_{\perp}}=&\sideset{}{''}\sum_{ijll'\sigma}t^{\perp}_{ij}\hat{c}^{\dagger}_{il\sigma}\hat{c}_{jl'\sigma}
+\sideset{}{''}\sum_{ ijll'}J^{\perp}_{ij}\hat{\mathbf{S}}_{il}\cdot\hat{\mathbf{S}}_{jl'}\\
+&U''\sideset{}{''}\sum_{ijll'}\big(\hat{c}^{\dagger}_{il\uparrow}\hat{c}^{\dagger}_{jl\downarrow}\hat{c}_{jl'\downarrow}\hat{c}_{il'\uparrow}
+\hat{c}^{\dagger}_{il\uparrow}\hat{c}^{\dagger}_{jl\downarrow}\hat{c}_{il'\downarrow}\hat{c}_{jl'\uparrow}\big)\\
+&U'\sideset{}{''}\sum_{ill'}\hat{c}^{\dagger}_{il\uparrow}\hat{c}^{\dagger}_{il\downarrow}\hat{c}_{il'\downarrow}\hat{c}_{il'\uparrow},
\end{split}
 \label{eq:H_start_perp}
\end{equation}
describes a number of possible interlayer dynamical processes. The primed (double primed) summation means that $i\neq j$ (and $l\neq l'$), where $i,j$ are indices of sites within the layer and $l,l'$ enumerate the layers. The first two terms of the intralayer 
Hamiltonian (\ref{eq:H_start_parallel}) represent the Hubbard model which consist of the hopping and the intrasite Coulomb repulsion terms.
The third term represents the antiferromagnetic exchange interaction. We take into account the nearest and the next-nearest
neighbor hopping terms with $t=-0.35$ eV and $t'=0.25|t|$, respectively (unless stated otherwise). Moreover, the $J_{ij}$ integrals are
assumed nonzero only for the nearest neighbors, for which $J_{ij}\equiv J\equiv 0.25|t|$. The intraatomic Coulomb repulsion is set to $U=14|t|$, unless stated otherwise. The model can thus be considered as either the extended Hubbard or $t$-$J$ Hamiltonian with $J$ originating from $p$-$d$ superexchange and a moderately high value of $U$ in these charge-transfer Mott-Hubbard systems\cite{Spalek2016}. 

The first two terms of the interlayer Hamiltonian $\hat{\mathcal{H}}_{\perp}$ represent the 
interlayer hopping and exchange, respectively. The hopping parameters $t^{\perp}_{ij}$
are chosen so that in reciprocal space the single-particle part takes the form 
\begin{equation}
 \mathcal{H}^t_{\perp}=\sideset{}{''}\sum_{\mathbf{k}ll'\sigma}\bigg[t_{\textrm{bs}}+\frac{t_z}{4}(\cos k_x - \cos k_y)^2 \bigg]\;\hat{c}^{\dagger}_{\mathbf{k}l\sigma}\hat{c}_{\mathbf{k}l'\sigma}\;,
\label{eq:int_hop_k}
\end{equation}
where $t_{\textrm{bs}}$ introduces the $\mathbf{k}$-independent bilayer splitting, whereas the $\mathbf{k}$-dependent part is taken in the usually form \cite{Andersen1995,Andersen1994} and is responsible for a highly anisotropic character of the splitting.
In our analysis, we assume that $J^{\perp}_{ij}$ is nonzero for $i=j$ only. The remaining two terms of (\ref{eq:H_start_perp}) represent the
so-called on-site and off-site interlayer pair hoppings \cite{Nishiguchi2013}, respectively. They originate from the long-range 
Coulomb interaction. Follwing Ref. \onlinecite{Nishiguchi2013} we set $U'=-2U''$. In our analysis we take into account only the nearest-neighbor interlayer pair hoppings (for $|\mathbf{R}_i-\mathbf{R}_j|=1$). Nevertheless, the interlayer pair hopping generates four-site terms (two sites on the first layer and two on the second) and it
is very difficult to take them into account within the DE-GWF method in higher orders. Hence, their influence
is going to be taken into account within the zeroth-order calculation scheme, which is equivalent to the modified Gutzwiller
approximation (SGA)\cite{Jedrak2011,Kaczmarczyk2011,Zegrodnik2014}. 
Such an approach is justified by the fact the $U'$ and $U''$ interaction parameters do not represent the 
dominant energies in the system (i.e., they are much smaller than $U$).

We have checked that the interlayer density-density Coulomb interaction term, $\sideset{}{''}\sum_{ill'}V_{ll'}\hat{n}_{il}\hat{n}_{il'}$, has a very small negative influence on the paired state. That is why we do not include this term in the analysis presented below. The influence of the intralayer intersite Coulomb repulsion is also omitted here; its influence on the SC state has been discussed by us separately within the single-layer model in Ref. \onlinecite{Abram2016}.

A methodological remark is in place here. Namely, in principle the last term in (\ref{eq:H_start_perp}) should also have a corresponding term in (\ref{eq:H_start_parallel}) $\sim J\sideset{}{'}\sum_{ijl}\hat{c}^{\dagger}_{il\uparrow}\hat{c}^{\dagger}_{il\downarrow}\hat{c}_{jl\downarrow}\hat{c}_{jl\uparrow}$. However, it is of the order $Jd^2$, where $d^2=\langle \hat{n}_{i\uparrow}\hat{n}_{i\downarrow} \rangle$ is the double occupancy probability, and under the circumstance that $J\ll U$, as well as $J$ is substantially smaller than $|U'|$ it can be safely neglected. Note also that the present formulation is based on real-space pairing, whereas the original formulation was based on a direct Josephson tunneling of the Cooper pairs in $\mathbf{k}$-space\cite{Chakravarty1993,Anderson1994, Chen2012_2}. The coherent  pair tunneling across the planes has also been confirmed\cite{Hettinger1995}. 

\section{Method}

To take into account the strong electronic correlations we start with the Gutzwiller-type wave function in the form
\begin{equation}
 |\Psi_G\rangle\equiv\hat{P}|\Psi_0\rangle=\prod_{il}\hat{P}_{il}|\Psi_0\rangle \;,
\end{equation}
where $|\Psi_0\rangle$ represents an uncorrelated state (chosen as the superconducting uncorrelated state) and
\begin{equation}
 \hat{P}_{il}\equiv \sum_{\Gamma}\lambda_{il}|\Gamma\rangle_{il\;il}\langle\Gamma|\;,
 \label{eq:P_Gamma}
\end{equation}
with $\lambda_{il}$ being the variational parameters and $|\Gamma\rangle_{il}$ representing the states from the
local basis
\begin{equation}
|\Gamma\rangle_{il}\in \{|\varnothing\rangle_{il}, |\uparrow\rangle_{il}, |\downarrow\rangle_{il},
|\uparrow\downarrow\rangle_{il}\}\;.
\end{equation}
The consecutive states represent empty, singly, and doubly occupied local configurations, respectively.
To significantly simplify the calculations and improve the convergence, one customarily imposes the condition\cite{Bunemann2012,Gebhard1990} 
\begin{equation}
 \hat{P}_{il}^2\equiv 1+x_{il}\hat{d}^{\textrm{HF}}_{il}\;,
 \label{eq:condition}
\end{equation}
where $x_{il}$ is a variational parameter and $\hat{d}^{\textrm{HF}}_{il}=\hat{n}_{il\uparrow}^{\textrm{HF}}\hat{n}_{il\downarrow}^{\textrm{HF}}$, $\hat{n}_{il\sigma}^{\textrm{HF}}=\hat{n}_{il\sigma}-n_{l0}$, 
with $n_{l0}=\langle\Psi_0|\hat{n}_{il\sigma}|\Psi_0\rangle$. By using Eqs. (\ref{eq:P_Gamma}) and (\ref{eq:condition}),
one can express the parameters $\lambda_{il}$ with the use of $x_{il}$. Furthermore, by assuming spatial 
homogeneity within the layer we are left with the two variational parameters $x_l$ ($l=1,2$). For the case with two equivalent layers, which is analyzed in detail here, we can additionally set $x_1=x_2$. However, at the end of the paper we also consider the case with different interaction parameters for the two layers. The approach allows
to express the expectation values, in the correlated state $|\Psi_G\rangle$, corresponding to each of the terms in the starting Hamiltonian (\ref{eq:H_start_general}). For example, for the case of the hopping term the corresponding expectation value takes the form
 \begin{equation}
 \begin{split}
  \langle\Psi_G|&\hat{c}^{\dagger}_{il\sigma}\hat{c}_{jl'\sigma}|\Psi_G\rangle=\\
  &\sum_{k=0}^{\infty}\frac{1}{k!}\sideset{}{'}\sum_{m_1f_1...m_kf_k}x^{k_1}_1x^{k_2}_2
  \langle\tilde{c}^{\dagger}_{il\sigma}\tilde{c}_{jl'\sigma}\hat{d}^{\textrm{HF}}_{m_1f_1}...\hat{d}^{\textrm{HF}}_{m_kf_k} \rangle_0\;,
  \end{split}
  \label{eq:diag_sum}
 \end{equation}
where $\hat{d}^{\textrm{HF}}_{\varnothing}\equiv 1$, $\tilde{c}^{(\dagger)}_{il\sigma}\equiv \hat{P}_{il}\hat{c}^{(\dagger)}_{il\sigma}\hat{P}_{il}$, 
and the index $m$ corresponds to lattice sites within the layer, whereas $f$ labels the layers. The primmed summation on the right hand side is restricted to $(l_p,m_p)\neq (l_{p'},m_{p'})$, $(l_p,m_p)\neq (i,l)$, $(l_p,m_p)\neq (j,l')$ for all $p$, $p'$.
The powers indices $k_1$ ($k_2$) express how many times the indices $f_p$ on the right hand of the equation
have the value $1$ ($2$) for a given term. They fulfill the relation $k_1+k_2=k$. 
The maximal $k$, for which the terms in Eq. (\ref{eq:diag_sum}) are taken into account, represents the order of the expansion.

By using the Wicks theorem in real space one can express all the averages in the noncorrelated state 
($\langle...\rangle_0\equiv\langle\Psi_0|...|\Psi_0\rangle$), which appear in the 
formulas for the ground state energy of the system in the form of diagrams, with vertices attached to the lattice sites and the edges corresponding to the so-called paramagnetic lines, $P_{ijll'\sigma}\equiv\langle\hat{c}^{\dagger}_{il\sigma}\hat{c}_{jl'\sigma} \rangle_0$,
or the superconducting lines, $S_{ijll'}\equiv\langle\hat{c}^{\dagger}_{il\uparrow}\hat{c}^{\dagger}_{jl'\downarrow} \rangle_0$. 
In our calculations we take into account lines for which $|\Delta \mathbf{R}|=|\mathbf{R}_{il}-\mathbf{R}_{jl'}|\leq 3$. 

When it comes to the SC lines we assume the $d$-$wave$ spin-singlet pairing with no intrasite pairing components, which corresponds to the strong correlation limit for the copper-based superconductors. The approach has been discussed extensively before \cite{Wysokinski2015,Wysokinski2016,Kaczmarczyk2013,Kaczmarczyk2014,Spalek2016}. For the interlayer pairing contribution we take the $s^{\pm}$ pairing symmetry\cite{Maier2011,Ubbens1994,Lee1995}. 

The method described briefly above allows to express the expectation value of our Hamilotnian in the correlated state
\begin{equation}
 \langle \hat{H}\rangle_G=\frac{\langle\Psi_G|\hat{H}|\Psi_G\rangle}{\langle\Psi_G|\Psi_G\rangle},
\end{equation}
as a function of all the lines $P_{ijll'\sigma}$, $S_{ijll'}$ and the variational parameters $x_1$, $x_2$.

To calculate the values of the paramagnetic and superconducting lines, as well as the variational parameters, the 
grand-canonial potential $\mathcal{F}=\langle \hat{H}\rangle_G-\mu_Gn_G$ is minimized, where $\mu_G$ and $n_G$ are respectively
the chemical potential and the number of particles per lattice site determined in the correlated state. The minimization condition together with the normalization of $|\Psi_0\rangle$, can be cast into the form of an effective Schr\"odinger equation
\begin{equation}
 \hat{H}_{\textrm{eff}}|\Psi_0\rangle=E_{\textrm{eff}}|\Psi_0\rangle,
 \label{eq:effective_Schrodinger}
\end{equation}
with the following effective Hamiltonian
\begin{equation}
 \hat{\mathcal{H}}_{\textrm{eff}}=\sum_{ijll'\sigma}t^{\textrm{eff}}_{ijll'}\hat{c}^{\dagger}_{il\sigma}\hat{c}_{jl'\sigma}
 +\sum_{ijll'}\big(\Delta^{\textrm{eff}}_{ijll'}\hat{c}^{\dagger}_{il\uparrow}\hat{c}^{\dagger}_{jl'\downarrow}+H.c.\big),
 \label{eq:H_effective}
\end{equation}
where the effective hopping and the effective superconducting gap parameters are defined through the relations 
\begin{equation}
 t^{\textrm{eff}}_{ijll'}\equiv \frac{\partial\mathcal{F}}{\partial P_{ijll'\sigma}},
 \quad \Delta^{\textrm{eff}}_{ijll'}\equiv \frac{\partial\mathcal{F}}{\partial S_{ijll'}}.
 \label{eq:effective_param}
\end{equation}
Such an approach has been commonly used in the literature \cite{Kaczmarczyk2013,Kaczmarczyk2014,Bunemann2012,Wang2006} and as shown in Ref. \onlinecite{Kaczmarczyk2014phmag}, it is equivalent to the method based on the Lagrange multipliers which in turn guarantees the preservation of the statistical consistency during the minimization procedure.

As one can see from (\ref{eq:effective_Schrodinger}) the uncorrelated state $|\Psi_0\rangle$ is an eigenstate of $\hat{H}_{\textrm{eff}}$, which has a BCS-like form with the kinetic energy term and the pairing term leading to nonzero expectation values $\langle\hat{c}^{\dagger}_{il\uparrow}\hat{c}^{\dagger}_{jl'\downarrow} \rangle_0$. Nevertheless, in contrast to the BCS theory, here, the pairing is defined in real space and it is not due to the phonon-mediated interaction. Instead, it stems from the inter-electronic correlations taken into account in higher orders via diagrammatic expansion. The terms for $l'\neq l$ in the second summation of (\ref{eq:H_effective}) corresponds to the interlayer pairing.

To determine $|\Psi_0\rangle$ we first transform the effective Hamiltonian to reciprocal space and then diagnonalize it by using the Nambu representation. This allows for the derivation
of the self-consistent equations for the superconducting and paramagnetic lines, as well as the chemical potential. Additionally, the parameters $\{x_l\}$ are determined variationally. 
After the determination of $P_{ijll'\sigma}$, $S_{ijll'}$, $\mu$, $x_1$, $x_2$, we can calculate
the correlated equivalents of the superconducting lines, 
$\langle\hat{c}^{\dagger}_{il\uparrow}\hat{c}^{\dagger}_{jl'\downarrow}\rangle_G$, which are regarded as the true gap parameters in the correlated state. In general, each non-correlated gap parameter
$\langle\hat{c}^{\dagger}_{il\uparrow}\hat{c}^{\dagger}_{jl'\downarrow} \rangle_0$ taken into account within the diagrammatic procedure has its correlated correspondent $\langle\hat{c}^{\dagger}_{il\uparrow}\hat{c}^{\dagger}_{jl'\downarrow} \rangle_G$.
The symmetry of the correlated gap reflects the symmetry assumed for the non-correlated gap (intralayer $d$-$wave$ and interlayer $s^{\pm}$ wave).
As it has been pointed out in Refs. \onlinecite{Kaczmarczyk2013}, \onlinecite{Kaczmarczyk2014}, \onlinecite{Abram2016}, the dominant contribution to the superconducting phase comes from $\langle\hat{c}^{\dagger}_{il\uparrow}\hat{c}^{\dagger}_{jl\downarrow}\rangle_G$ with $\mathbf{R}_i-\mathbf{R}_j=(a,0)$ for the $d$-$wave$ intralayer pairing. For the $s^{\pm}$ interlayer pairing the dominant contribution comes from $\langle\hat{c}^{\dagger}_{il\uparrow}\hat{c}^{\dagger}_{jl'\downarrow}\rangle_G$ for $\mathbf{R}_i-\mathbf{R}_j=(0,0)$ and $l\neq l'$. Those two intra- and inter-layer correlated gap parameters are going to be denoted by $\Delta^{||}_G$ and $\Delta^{\perp}_G$ throughout the paper and lead to the following $\mathbf{k}$-dependence of the overall gap
\begin{equation}
 \Delta_{G}(\mathbf{k})=2\Delta^{||}_G(\cos k_x + \cos k_y) \pm \Delta^{\perp}_G,
\end{equation}
where the $\pm$ sign corresponds to bonding and antibonding hybridized bands, respectively.


\section{Results and discussion}
Our analysis is performed in the fourth order of the diagrammatic expansion, if not stated otherwise. As shown in Refs. \onlinecite{Kaczmarczyk2014} and \onlinecite{Abram2016}, for the considered types of models, satisfactory convergence is achieved in the 4-5 order of the expansion. We focus mainly on the $t$-$J$-$U$ bilayer case analysis (Subsection A). As it has been already stated we set the intralayer model parameters to $t=-0.35$ eV, $t'=0.25|t|$, $U=14|t|$, $J=0.25|t|$, unless stated otherwise. In the following we provide the values of the model parameters in the units of $|t|$. For the sake of completeness we also present selected results for the case of the Hubbard model (Subsection B).

\subsection{Results for the $t$-$J$-$U$ model}

\begin{figure}[!h]
\centering
\epsfxsize=80mm 
{\epsfbox[140 401 485 787]{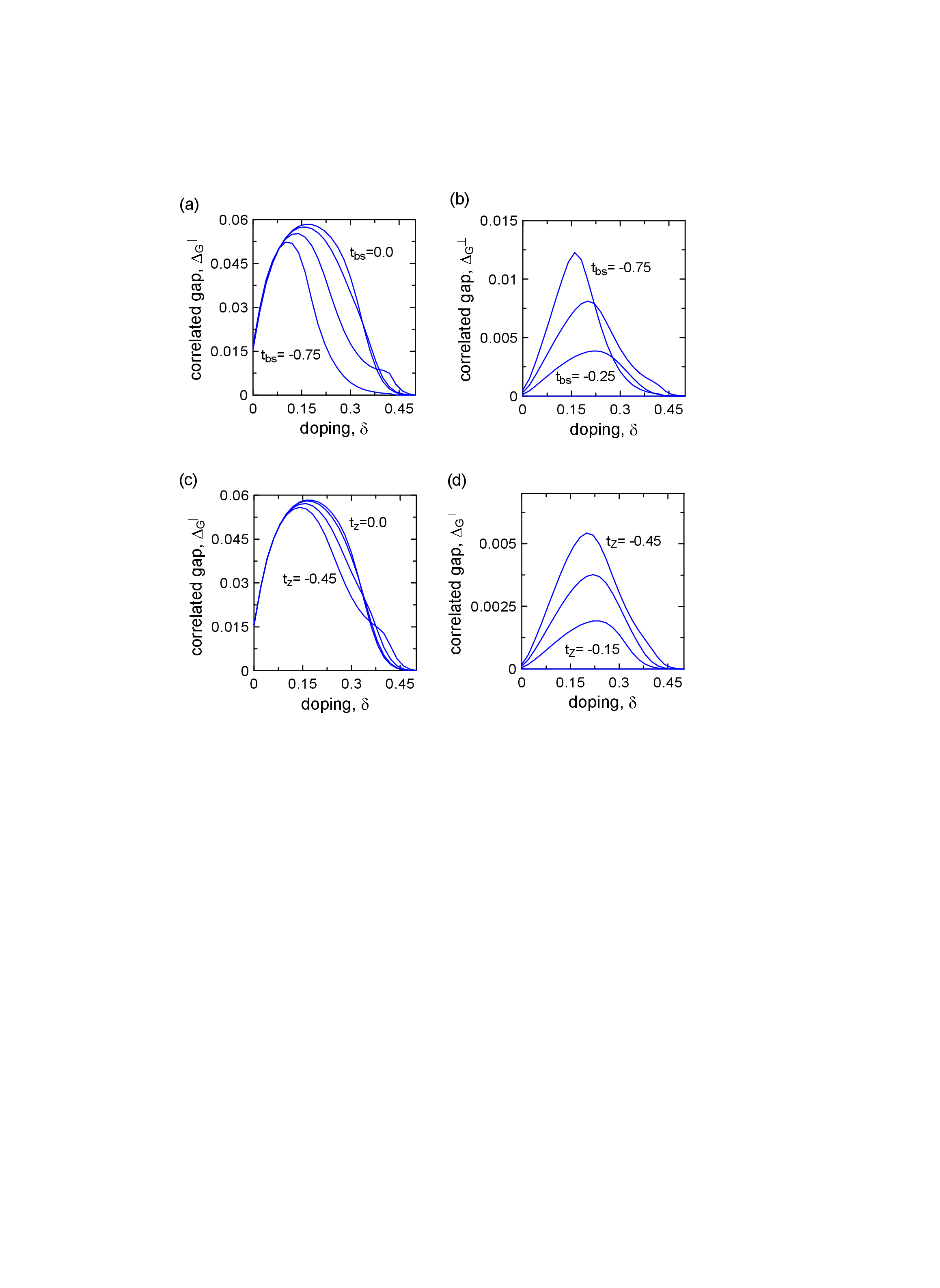}}
\caption{(a), (b) Dimensionless intra- ($\Delta^{||}_G$) and inter-layer ($\Delta^{\perp}_G$) correlated gap parameters, both as a function of doping for $t_z=0$, and $t_{\textrm{bs}}$ varying between 0.0 and -0.75, with the step -0.25. For $t_{\textrm{bs}}=0$ the interlayer gap is zero in the whole doping range. (c), (d) the intra- and inter-layer correlated gap parameters, respectively, as a function of doping for $t_{\textrm{bs}}=0$ and $t_z$ varying between 0.0 and -0.45 with the step -0.15. Similarly as above, the interlayer gap is zero for the case with $t_z=0$. For nonzero $t_z$ both gaps vanish at the same doping. The results correspond to $J^{\perp}=0$.}
\label{fig:1}
\end{figure}

First, we analyze the influence of the interlayer electron hopping terms on the paired state, with no interlayer exchange interaction included ($J^{\perp}=0$). The magnitude of the $\mathbf{k}^{||}$-dependent part of the interlayer hopping is governed by $t_z$ (cf. Eq. (\ref{eq:int_hop_k})) and is responsible for the anisotropy of the bilayer splitting. This term leads to a significant splitting in the antinodal direction ($\pi$,$0$) and no splitting in the nodal one ($\pi$,$\pi$). In some papers referring to the bilayer cuprate superconductors, the nodal bilayer splitting is neglected, i.e., the situation with $t_{bs}=0$ is considered \cite{Mori2006,Nishiguchi2013,Chen2012}. At the same time, considerations with the nearest-neighbor interlayer hopping included only (with $t_z=0$) have been also carried out \cite{Maier2011,Lanata2009,Medhi2009,Voo2015}. In the latter situation the splitting between the bonding and the antibonding bands is isotropic which is rather loosely connected to the 
electronic 
structure of the copper-based compounds. In Fig. \ref{fig:1} we show the influence of both the isotropic and anisotropic 
bilayer splittings on the paired state. As one can see, the effect of increasing either $t_{\textrm{bs}}$ or $t_z$ is similar. Due to the hopping of electrons between the layers the intralayer correlated gap diminishes mainly in the overdoped regime, whereas the interlayer one increases. This is caused by the fact that if there is no interlayer interaction ($J^{\perp}=0$), the hopping mainly dilutes the intralayer pair bonds and thus weakens the pairing. Nevertheless, in the underdoped regime the gap parameter $\Delta^{||}_G$ remains almost unchanged by the interlayer hopping. Also, the interlayer gap is approximately one order of magnitude smaller than the intralayer one. As one would expect, the case with $t_z=t_{\textrm{bs}}\equiv 0$ leads to intralayer pairing only ($\Delta^{\perp}_G=0$). In the zeroth order of the DE-GWF calculations, which are equivalent to the RMFT approach, one obtains no interlayer pairing even in the case of nonzero interlayer hopping.

\begin{figure}[!h]
\centering
\epsfxsize=80mm 
{\epsfbox[140 549 470 728]{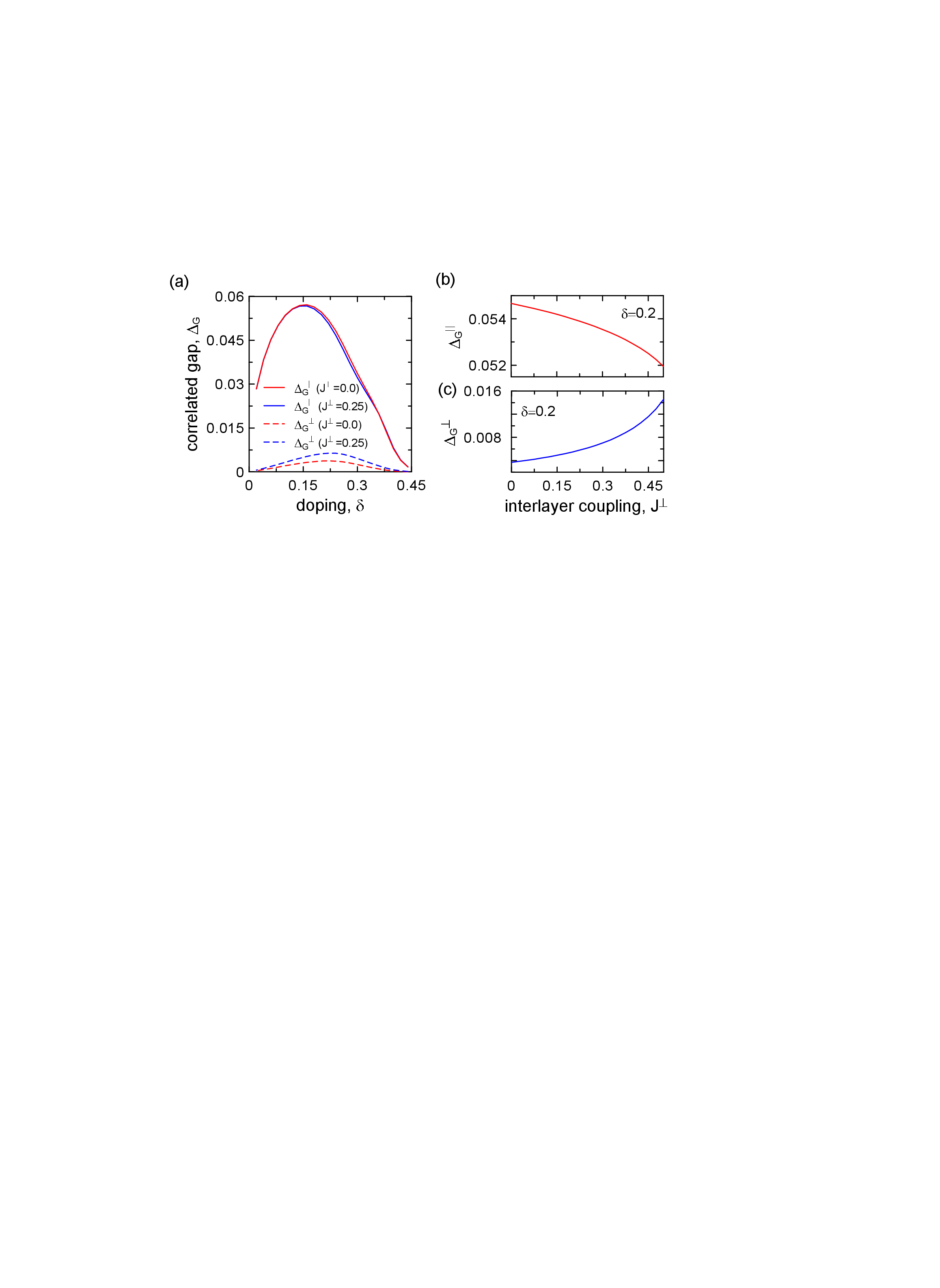}}
\caption{(a) Intra- and inter-layer correlated gap parameters as a function of doping for two selected values of $J^{\perp}$, for $t_z=-0.3$ and $t_{\textrm{bs}}=0$. (b) and (c) the intra- and inter-layer correlated gap parameters, respectively, as a function of the interlayer exchange coupling for the selected value of doping $\delta=0.2$ which is close to optimal doping. The effect of $J^{\perp}$ is minor and thus the principal single-plane features of the SC state are preserved.}
\label{fig:2}
\end{figure}

Next, we scrutinize the influence of the interlayer exchange coupling $J^{\perp}$ on the SC gap. The results are presented in Fig. \ref{fig:2}. It can be seen that the $J^{\perp}$ term has a very small influence on the intralayer pairing while its interlayer correspondent is significantly increased (almost twice) when changing the value of $J^{\perp}$ from 0.0 to 0.25. It should be noted that even for the case when $J^{\perp}=J$ (i.e., when the intra- and inter-layer exchange couplings are equal), the intralayer gap parameter is still about one order of magnitude larger that the interlayer one. This effect should be related to the fact that the system is infinite in the ($x$,$y$) plane whereas it is limited in the $z$-direction (two layers only).

\begin{figure}[!h]
\centering
\epsfxsize=80mm 
{\epsfbox[136 429 465 610]{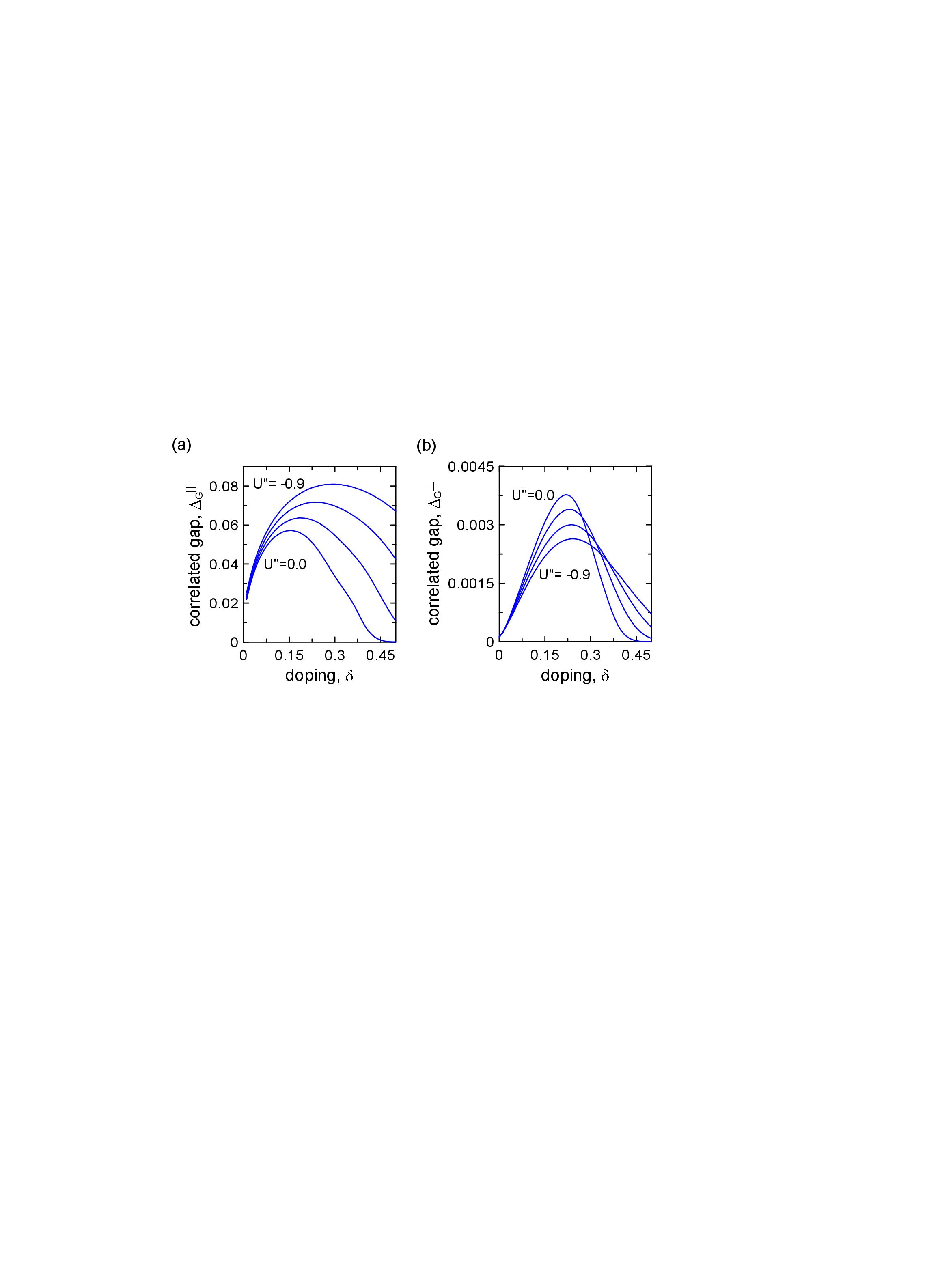}}
\caption{Intralayer (a) and interlayer (b) correlated gaps as a function of doping for $t_z=-0.3$, $t_{\textrm{bs}}=0$, $J^{\perp}=0$ and for selected values of the pair hopping tunelling magnitude $U''$ varying between $0.0$ and $-0.9$ with the step $-0.3$, assuming $U'=-2U''$. The substantial value of $U''\leqslant -0.3|t|$ does not reporoduce the experimental phase diagram with $\Delta_G=0$ for $\delta\gtrsim 0.4$. }
\label{fig:Uph}
\end{figure}

In Fig. \ref{fig:Uph} we also show the influence of the interlayer pair hopping processes on the superconducting gap. Following Nishiguchi et al. \cite{Nishiguchi2013}, we set $U'=-2U''$. As one could expect, the inclusion of the pair tunneling process enhances the intralayer SC gap and broadens the SC stability regime. However, due to the mentioned effect the maximum value of the interlayer gap is decreased. The positive effect on the interlayer correlated gap is significant even though the values of the $U"$ parameter have been taken as relatively small in comparison to the Hubbard $U$. Note that the interlayer pair hopping term within this analysis originates from the long-range Coulomb interaction\cite{Nishiguchi2013}.

From Figs. \ref{fig:1}-\ref{fig:Uph} one can see that the most important factor in changing effectively the gap is the Coulomb interactions ($U'$, $U''$) leading to the interlayer pair hopping processes. The role of the number of layers in the elementary cell in increasing the stability of the SC phase was analyzed earlier within the mean-field approach\cite{Byczuk1995}.

\begin{figure}[!h]
\centering
\epsfxsize=90mm 
{\epsfbox[202 280 545 726]{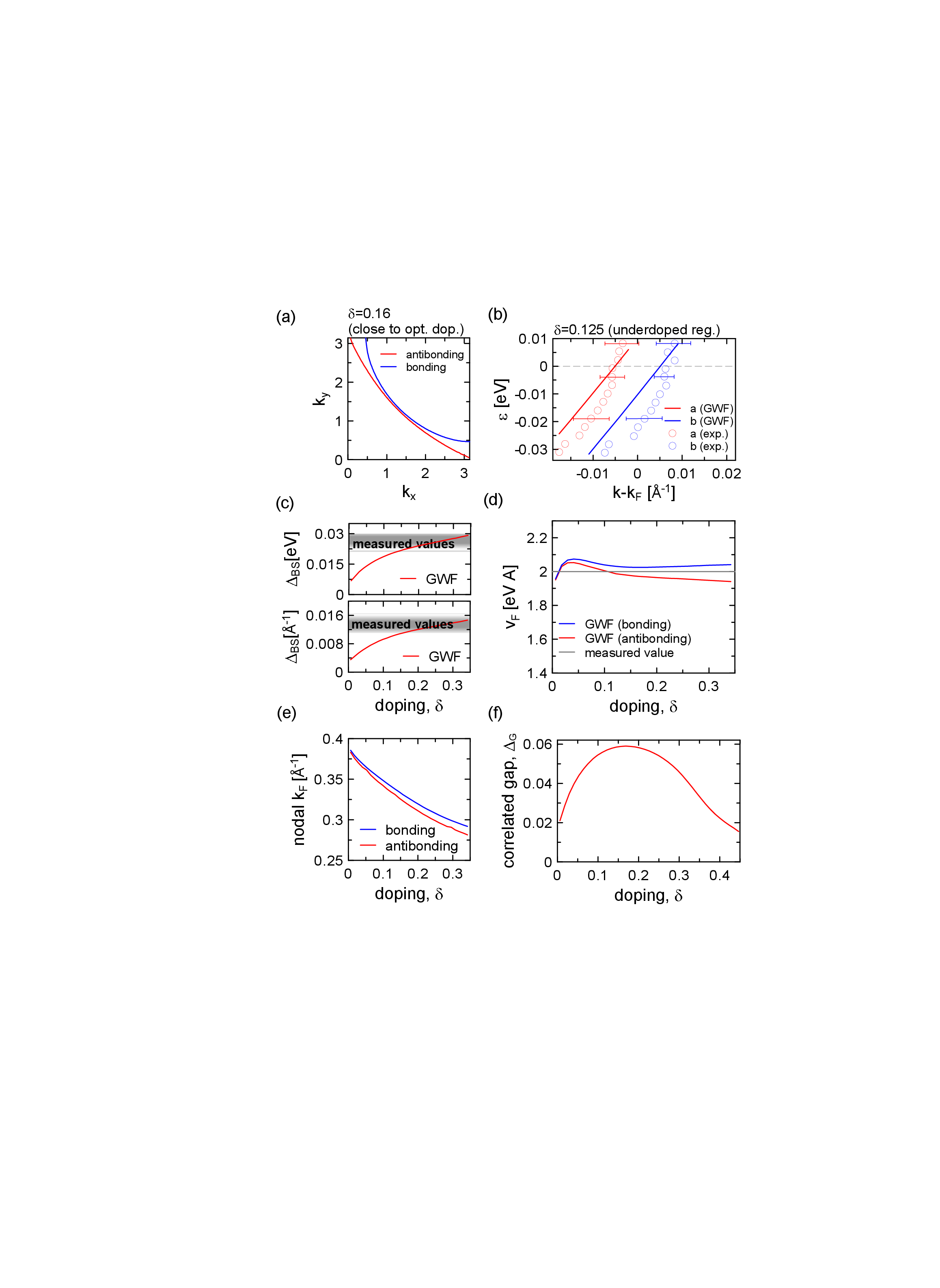}}
\caption{(a) Fermi surface sheets close to the optimal doping. (b) Theoretical (solid lines) and experimental (circles) dispersion relations in the underdoped regime. The experimental data have been taken from Ref. \onlinecite{Kordyuk2004}. (c) The nodal bilayer splitting in energy (top) and in the reciprocal space (bottom), both as a function of doping. The shaded regions correspond roughly to the measured values reported in Ref. \onlinecite{Kordyuk2004}. (d) Nodal Fermi velocity as a function of doping corresponding to the bonding and antibonding bands. The gray (solid) line corresponds to the measured value taken from Ref. \onlinecite{Kordyuk2004} which refer to the bonding band. (e) Fermi momenta corresponding to the bonding and antibonding bands, both as a function of doping. (f) Intralayer correlated gap magnitude as a function of doping. The presented fittings have been obtained for the following values of the model parameters: $t'=0.36$, $t_z=-0.15$, $t_{\textrm{bs}}=-0.06$, $U=13.5$, $J=0.25$ (all 
in units of $|t|$).}
\label{fig:4}
\end{figure}

In the remaining part of this subsection we show the results for the case when the model parameters are set to reproduce the selected principal characteristics measured by ARPES experiments of Ref. \onlinecite{Kordyuk2004}. According to the experimental data obtained for Bi-2212, the bilayer splitting of the Fermi surface is highly anisotropic, with maximal value in the antinodal direction and small, but nonzero value in the nodal direction. Moreover, even in the overdoped regime both the bonding and the antibonding parts of the Fermi surface appear to be hole-like. To reproduce the data within our model both $t_z$ and $t_{\textrm{bs}}$ have to be taken as nonzero ($t_z=-0.1$ and $t_{\textrm{bs}}=-0.06$). In these considerations we have neglected the appearance of the interlayer pairing mechanism. Such a choice is justified by the fact that the interlayer gap magnitude is one order of magnitude smaller than its intralayer corespondant (cf. Fig. \ref{fig:1}).
 In Fig. \ref{fig:4} (a) and (b) we plot the calculated two sheets of Fermi surface and the dispersion relations close to the corresponding Fermi momenta for selected values of doping close to the optimal doping and in the underdoped regime, respectively. The theoretical composed Fermi surface topology shown in Fig. \ref{fig:4} (a) is roughly similar to that obtained experimentally in Ref. \onlinecite{Kordyuk2004} (cf. Fig. 1. in that paper) and the theoretical lines which reflect the dispersion relations in Fig. \ref{fig:4} (b) are relatively close to the experimental points. However, it should be noted that the authors of Ref. \onlinecite{Kordyuk2004} provide only the critical temperature of the corresponding samples which allows to determine the corresponding doping regions only (underdoped or overdoped region). The specific doping levels of the samples are not provided explicitly and this is one of the reasons why our theoretical results may not reflect precisely those coming from the experiment. In Fig. 
\ref{fig:4} (c) we 
plot the bilayer splitting in energy (top) and in 
the reciprocal space (bottom) in the nodal direction, both as a function of doping. 
The gray region represents the measured values for different doping levels presented in Ref. \onlinecite{Kordyuk2004}. The experimentally measured very weak doping dependence of the nodal bilayer splitting value has been confirmed in Ref. \onlinecite{Borisenko2006} for YBCO. However, for this compound, the value of $\Delta_{\mathrm{BS}}$ was almost five times larger than the one measured for Bi-2212. As one can see, our theoretical results are close to the experimental ones only in the overdoped regime while in the underdoped regime the calculated bilayer splitting significantly decreases. It should be noted 
that Fournier et al. \cite{Furnier2010} have reported a complete vanishing of the bilayer splitting below the optimal doping for the YBCO samples. The inconsistence between the different sets of experimental data from the one side, and the theoretical results from the other requires a further analysis. 

\begin{figure}[!h]
\centering
\epsfxsize=90mm 
{\epsfbox[202 591 510 733]{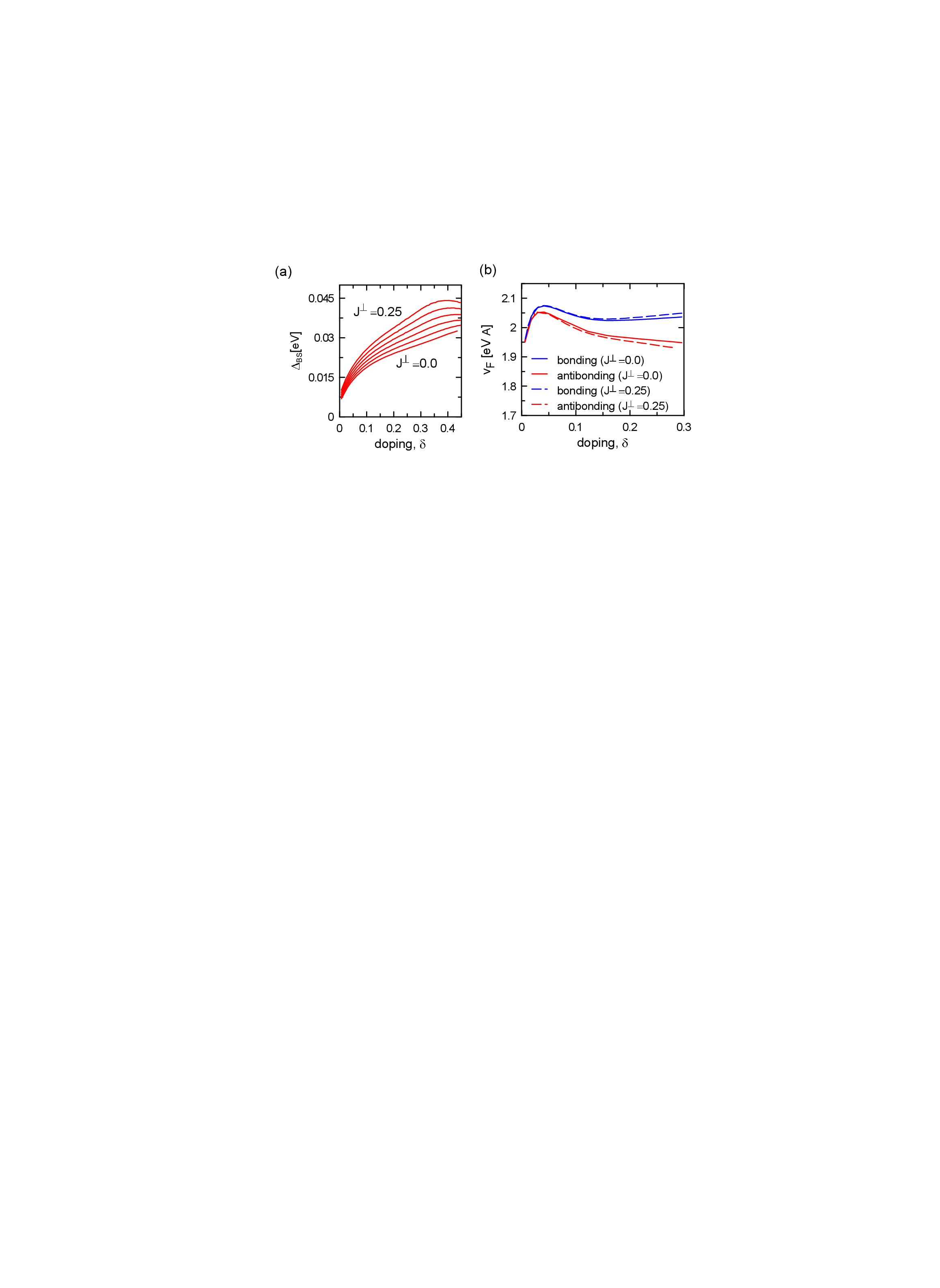}}
\caption{(a) The nodal bilayer splitting in energy as a function of doping for different values of the interlayer exchange coupling varying between $J^{\perp}=0.0$ and $J^{\perp}=0.25$ with the step $0.05$. (b) Nodal Fermi velocity as a function of doping corresponding to the bonding and antibonding bands for two values of the the interlayer exchange coupling.}
\label{fig:45}
\end{figure}

In Fig. \ref{fig:45} we show the effect of the interlayer coupling on the bilayer splitting and the nondal Fermi velocity. As one can see, with the increasing $J^{\perp}$ the bilayer splitting also increases. However, the Fermi velocity is very weakly dependent on the $J^{\perp}$ value and even for rather strong interlayer couplings the changes in $v_F$ are barely visible. The same situation takes place for the correlated gap which is also almost unaffected by this factor (cf. Fig. \ref{fig:2}).

Due to the nonzero splitting in the nodal direction, also two values of the nodal Fermi velocity appear, which are shown in Fig. \ref{fig:4} (d). The measured values reported in Ref. \onlinecite{Kordyuk2004} are close to $2.0$ eV\AA$\;$ and they correspond to the bonding band. Also, other experimental reports corresponding to the cuprate superconductors, demonstrate very weak doping dependence of the nodal Fermi velocity close to that value \cite{Zhou2003,Kordyuk2005,Borisenko2006}.
In this case, the weak doping dependence is reproduced by the theory as can be seen in Fig. \ref{fig:4} (d). Moreover, the bonding and the antibonding Fermi velocities are very close. Near half filling both of them have almost the same value. It is caused by the fact that the nodal bilayer splitting is decreasing as we are getting closer to the half filled situation, what is 
illustrated also in Fig. \ref{fig:4} (e). For the sake of completeness, in Fig. \ref{fig:4} (f) we also provide the correlated gap magnitude as a function of 
doping for the same set of model parameters as the remaining results presented in this Figure.

\subsection{Results for the Hubbard model}

For the sake of completeness, we have also analyzed the selected results for the case of the Hubbard model ($J\equiv 0$, $J^{\perp}\equiv 0$). For these calculations we set the interlayer hopping as non-dependent on $\mathbf{k}$ ($t_z=0$). In the main part of this Subsection we also neglect the interlayer pair hopping term ($U'=U''=0$). Nonetheless, the influence of nonzero $U'$ and $U''$ is shown briefly at the end.

\begin{figure}[!h] 
\centering
\epsfxsize=80mm 
{\epsfbox[196 594 510 779]{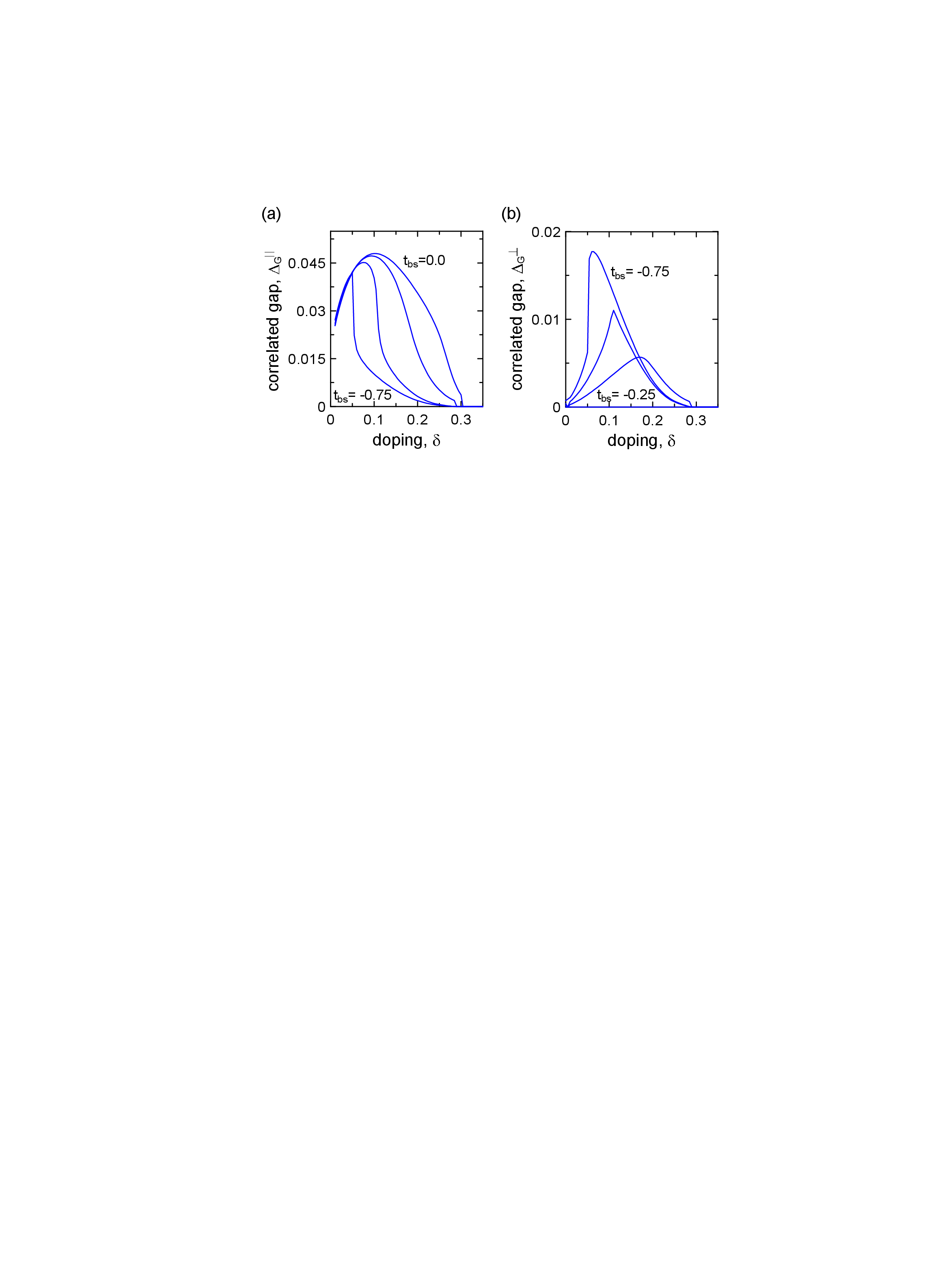}}
\caption{Intra- (a) and inter-layer (b) correlated gap vs. doping for $t_z=0$ for $t_{\textrm{bs}}$ varying between 0.0 and 0.75 with the step 0.25. For $t_{\textrm{bs}}=0$ the interlayer gap is zero in the whole doping range. The data correspond to the Hubbard model with $U=14$ and ($U'=U''=0$).}
\label{fig:3}
\end{figure}

As one can see in Fig. \ref{fig:3}, the effect of the splitting parameter $t_{\textrm{bs}}$ is similar as in the case of the $t$-$J$-$U$ model. However, for large values of $t_{\textrm{bs}}$, a discontinuity appears in the $\delta$-dependence of both $\Delta^{||}_G$ and $\Delta^{\perp}_G$ in the underdoped regime (the curves for $t_{\textrm{bs}}=0.75$), which has not been observed earlier. 

\begin{figure}[!h] 
\centering
\epsfxsize=90mm 
{\epsfbox[206 428 540 736]{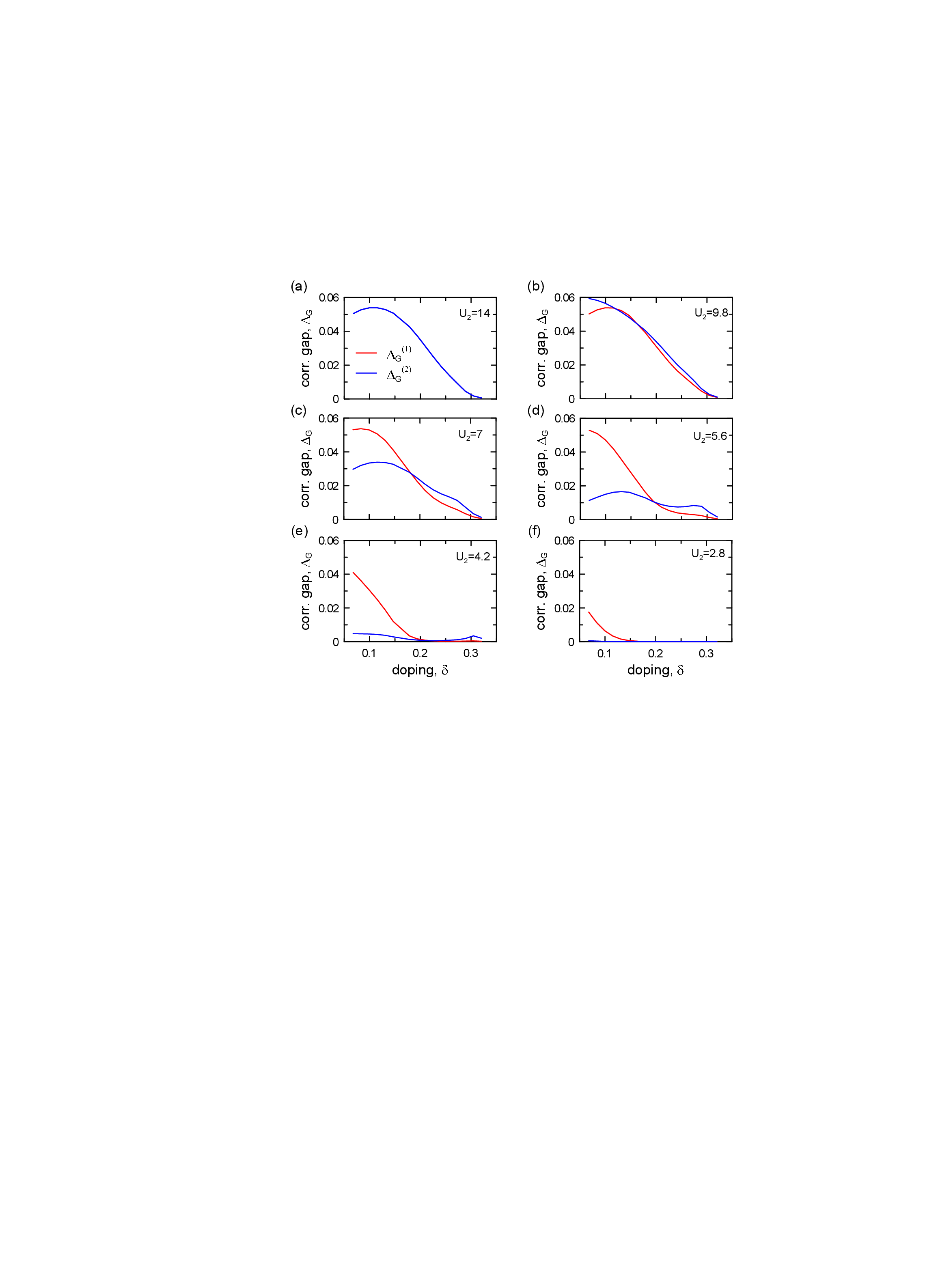}}
\caption{Correlated gaps for the bilayer Hubbard model in the first (red solid line) and second (blue solid line) layers vs. doping for different values of Coulomb repulsion in the second layer, $U_2$ (the explicit values are provided in the Figures). The value $U_1$ corresponding to the Coulomb repulsion in the first layer remains constant and equal to $U_1=14$ while the interlayer electron hopping parameters are set to $t_{\textrm{bs}}=-0.1$ and $t_z=0$. In (a) we have $U_1=U_2$ and both correlated gaps have the same values marked by blue line, while in (f) the value of $U_2$ is so low that the correlated gap in the second layer is practically zero. The data correspond to no interlayer pair hopping term ($U'=U''=0$).}
\label{fig:44}
\end{figure}

\begin{figure}[!h] 
\centering
\epsfxsize=85mm 
{\epsfbox[230 515 536 744]{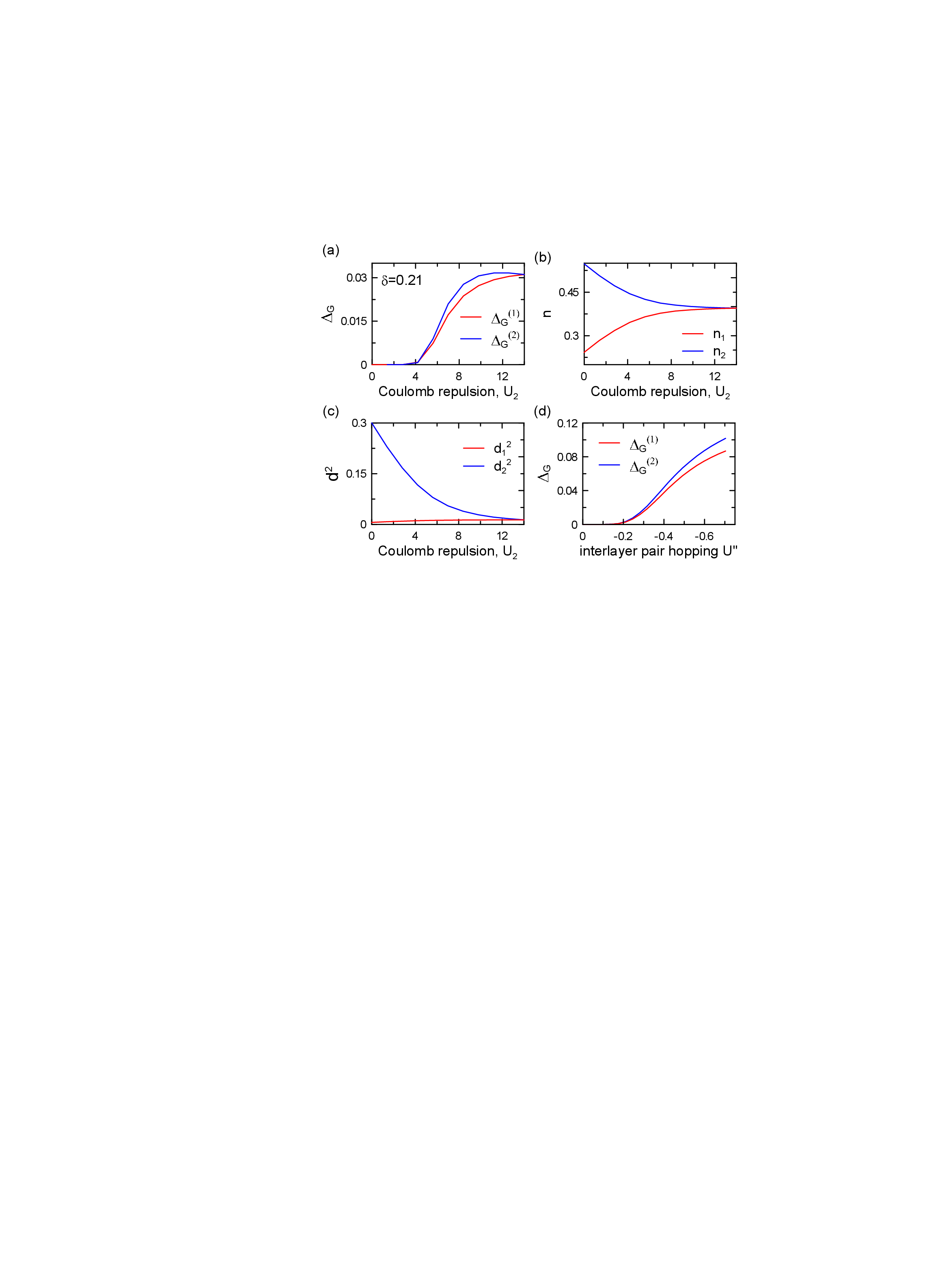}}
\caption{Intralayer correlated gap (a), the number of carriers per atomic site (b), and double occupancy probability (c) in the first (blue line) and second (red line) atomic layer, all as a function of Coulomb repulsion in the second layer. For (a), (b), and (c) the interlayer pair hopping is set to zero $U''=U'=0$. (d) Intralayer correlated gaps as functions of interlayer pair hopping for $U_2=0$. We keep $U'=-2U''$ fulfilled. For (a), (b), (c), and (d) the Coulomb repulsion in the first layer is set to $U_1=14$ and the interlayer electron hopping parameters are: $t_{\textrm{bs}}=-0.1$ and $t_z=0$, while the selected value of doping is $\delta=0.21$. }
\label{fig:46}
\end{figure}

We have also analyzed the influence of the disproportionation between the Coulomb repulsion in the two layers. In the following analysis $U_1$ and $U_2$ correspond to the Coulomb repulsion in the first and the second layer, respectively. In such a situation the intralayer correlated gaps, the average number of electrons per atomic site, and the average double occupancies in the two layers can have different values. All the mentioned quantities  are presented in Figs. \ref{fig:44} and \ref{fig:46}. It can be seen that the correlated gaps for both layers decrease as the difference between $U_1$ and $U_2$ increases. For $U_2=0$ both superconducting gaps vanish (cf. Fig. \ref{fig:46} (a)). In Fig. \ref{fig:46} (b) we show that in the considered system electrons are pushed from the layer with stronger Coulomb repulsion to that with the weaker. Obviously, the double occupancy in the second layer increases with decreasing $U_2$, as can be seen in Fig. \ref{fig:46} (c). For the case with $U_1=14$ and $U_2=0$ the 
paired phase 
vanishes, $\Delta_G^{(1)}=\Delta_G^{(2)}=0$ (cf. Fig. \ref{fig:46}(a)). However, after setting to interlayer pair hopping as nonzero, we can retain the stability of the paired phase in the system (cf. Fig. \ref{fig:46} (d)). In this situation, the pairing originating from the correlation effects in the layer with nonzero Coulomb repulsion is transfered to the second layer (with no Coulomb repulsion) by the pair hopping process. Such situation brings into mind the systems composed of one superconducting monolayer of correlated compound placed on a metallic layer in which the paired phase can be induced by the proximity effect \cite{Yuli2008}. However, in such interpretation the Cooper pair tunneling terms included in the Hamiltonian would have to originate from the Josephson coupling.

\section{Summary}

We start with a general remark. The majority of the papers concerning the real space pairing in high-Tc superconductors deal with a single Cu-O plane as reflecting the intrinsic properties of the copper-based compounds. In our recent (preceding) paper \cite{Spalek2016} we have shown that indeed in such an approach one can rationalize, even in a quantitative manner, the principal ground-state properties of high-T$_C$ superconductors. Nevertheless, the basic question still remains as to whether the model two-dimensionality, even if realistic, provides similar results when the basic aspects of three-dimensionality are incorporated into the theoretical scheme. Our present work provides an affirmative answer to this question, even though the detailed features are model-parameter dependent. We believe that the direct relation between theory and experiment analyzed here and in our previous work\cite{Spalek2016}, can contribute significantly to the resolution of the fundamental question whether a purely electronic 
model based on 
electronic correlations (without a glue\cite{Anderson2007}) can rationalize the superconducting properties of the copper systems. 

Turning to our results here, we have analyzed the influence of different interlayer dynamical processes on the superconducting state within the bilayer $t$-$J$-$U$ model. When it comes to the bilayer splitting, induced by the interlayer electron hopping, both the anisotropic ($\mathbf{k}^{||}$-dependent term with the magnitude tuned by parameter $t_z$) and the isotropic (of magnitude $t_{\textrm{bs}}$) contributions have been analyzed. Our analysis shows that both terms have a similar influence on the supercodncuting gap parameters (cf. Fig. \ref{fig:1}). With the increasing interlayer hopping magnitudes the intralayer SC gap decreases and that of the interlayer part increases. For the case of the bilayer Hubbard model, for large value of $t_{\textrm{bs}}=-0.75$ the discontinuity appears in the doping dependence of the SC gap (cf. Fig. \ref{fig:3}). As this effect has not been observed experimentally, the $t_{\mathrm{bs}}$ value must be essentially smaller than the value of $|t|$.

It has been reported recently in Ref. \onlinecite{Voo2015} that the results of the VMC calculations for the bilayer $t$-$J$-$U$ model for $d$-$wave$ superconducting phase are insensitive to $t_{\perp}$ in the range $|t_{\perp}|<0.5$ for $\delta <0.2$. As one can see from Fig. \ref{fig:1} also within our analysis both interlayer hopping terms do not influence significantly the correlated gap amplitude, $\Delta_G^{||}$, in the underdoped regime for not too large values of $|t_{\mathrm{bs}}|$ and $|t_{z}|$.

The main influence of the interlayer exchange coupling is seen in the interlayer superconducting gap magnitude, while the intralayer gap is only slightly reduced with increasing $J^{\perp}$ (cf. Fig. \ref{fig:2}). Weak dependence of the intralayer pairing on $J^{\perp}$ is also mentioned in Ref. \onlinecite{Voo2015}. The nonzero exchange integral $J^{\perp}$ explains the nature of antiferromagnetism in the insulating state\cite{Coldea2001,Chakravarty1989}. Therefore, it is tempting to conclude that for nonzero and small values of $J^{\perp}$ one can rationalize the nature of AF state and at the same time uphold the validity of the principal results concerning the superconducting phase obtained for the single-plane systems.

Among the considered interlayer processes the most significant influence on the SC phase has been reported for the interlayer pair hopping. The corresponding terms have also been included within the analysis of the bilayer Hubbard model in Ref. \onlinecite{Nishiguchi2013}. Similarly as in our case, it is necessary to include at least one of the off-site terms to achieve a meaningful enhancement of the SC phase stability. In our situation, this is a direct consequence of the fact that we assume $\langle\hat{c}^{\dagger}_{il\uparrow}\hat{c}^{\dagger}_{il\downarrow} \rangle\equiv 0$, as a result of the strong on-site Coulomb repulsion. Within the analysis shown in Ref. \onlinecite{Nishiguchi2013} the contribution to the pairing, resulting from the interplay between the 
three pair-hopping terms (one on-site and two off-site pair hoppings, cf. Eq. \ref{eq:H_start_perp}), seems not to be straightforward. However, their analysis is carried out within the spin-fluctuation picture for the case of weak interactions, what means that their approach is very different from ours.


We have also analyzed the basic features which refer to the available data obtained by ARPES experiments (such as the Fermi surface two-sheet structure, the bilayer splitting, the dispersion relations, the nodal Fermi velocity). In this part of our analysis, the model parameters have been fitted and roughly reproduce the experimental data presented in Ref. \onlinecite{Kordyuk2004}. As one can see, the highly anisotropic character of the bilayer splitting, with a small value of the splitting in the nodal direction, has been reproduced (cf. Fig. \ref{fig:4} (a)). The calculated value of the bilayer splitting is similar to that measured in the experiment for overdoped samples (cf. Fig. \ref{fig:4} (b)). However, in the underdoped regime significant discrepancies between our theory and experiment appear (cf. Fig. \ref{fig:4} (c) and (e)). The mentioned report\cite{Kordyuk2004} does not provide specific values of the doping levels, so the precise comparison cannot by carried out. At this point, 
it should be noted 
that in disagreement with Ref. \onlinecite{Kordyuk2004}, Fournier et al. \cite{Furnier2010} have reported a vanishing bilayer splitting below the optimal doping for the YBCO samples. Such a behavior is not reproduced within our approach. Finally, we also show that the nodal Fermi velocities, corresponding to the bonding and the antibonding bands, are very similar and close to the value measured experimentally ($\approx 2.0$ eV\AA$\;$) in the whole doping range (cf. Fig. \ref{fig:4} (d)). The present analysis for the bilayer $t$-$J$-$U$ model complements our recent results\cite{Spalek2016} for a single Cu-O planar superconductor. 

The results presented for the case of the bilayer Hubbard model with the two different values of the Coulomb repulsion for the two layers show that the disproportionation between $U_1$ and $U_2$ leads to different carrier concentrations in the layers, what in turn has a negative influence on the paired phase (cf. Fig. \ref{fig:46} a and b). However, for the situation representing a metallic monolayer in contact with a strongly correlated superconducting monolayer ($U_1=14$ and $U_2=0$), the paired phase can be induced by the Cooper pair tunneling processes (cf. Fig. \ref{fig:46} d).

The bilayer Hubbard model considered here might also be of interest in reference to the ultra-cold atom systems analyzed in recent years\cite{Nishida2010,Klawunn2010}.

\section{Acknowledgement}
We would like to acknowledge the financial support from the Grant MAESTRO, No. DEC-2012/04/A/ST3/00342 from the National Science Centre (NCN) of Poland.


\begin{thebibliography}{99}
\bibitem{Uchida2015}
Shin-ichi Uchida, \textit{High Temperature Superconductivity}, Springer Japan, Tokyo, 2015.

\bibitem{Kleiner1992}
R. Kleiner, F. Steinmeyer, G. Kunkel, and P. M\"uller, Phys. Rev. Lett. {\bf 68}, 2394 (1992).

\bibitem{Kordyuk2004}
A. A. Kordyuk et al., Phys. Rev. B {\bf 70}, 214525 (2004).

\bibitem{Borisenko2006}
S. V. Borisenko et al., Phys. Rev. Lett. {\bf 96}, 117004 (2006).

\bibitem{Garcia2010}
D. Garcia-Aldea and S. Chakravarty, New J. Phys. {\bf 12}, 105005 (2010).

\bibitem{Andersen1995}
O. K. Andersen, A. I. Lichtenstein, O. Jepsen, and F. Paulsen, J. Phys. Chem. Solids {\bf 56}, 1573 (1995).

\bibitem{Harrison2015}
N. Harrison, B. J. Ramshaw, and A. Shekhter Sci. Rep. {\bf 5}, 10914 (2015).

\bibitem{Medhi2009}
A. Medhi, S. Basu, and C. Kadolkar, Eur. Phys. J. B {\bf 72}, 583 (2009).

\bibitem{Maier2011}
T. A. Maier and D. J. Scalpino Phys. Rev. B {\bf 84}, 180513(R) (2011).

\bibitem{Lanata2009}
N. Lanata, P. Barone, and M. Fabrizio, Phys. Rev. B {\bf 80}, 224524 (2009).

\bibitem{Mori2006}
M. Mori, T. Tohyama, and S. Maekawa, J. Phys. Soc. Jpn. {\bf 75}, 034708 (2006).

\bibitem{Nishiguchi2013}
K. Nishiguchi, K. Kuroki, R. Arita, T. Oka, and H. Aoki, Phys. Rev. B {\bf 88}, 014509 (2013).

\bibitem{Voo2015}
K.-K. Voo, Phys. Lett. A {\bf 379}, 1743 (2015).

\bibitem{Ruger}
R. R\"uger, L. F. Tocchio, R. Valenti, and C. Gros, New J. Phys. {\bf 16}, 033010 (2014).

\bibitem{Medhi2007}
A. Medhi, S. Basu, C. Y. Kadolkar, Phys. Rev. B {\bf 76}, 235122 (2007).

\bibitem{Medhi2007_2}
A. Medhi, S. Basu, C. Y. Kadolkar, Physica C {\bf 451}, 13 (2007).

\bibitem{Zhao2007}
G. Zhao, Phys. Rev. B {\bf 75}, 140510(R) (2007).

\bibitem{Ding1995}
H. Ding et al., Phys. Rev. Lett. {\bf 74}, 2784 (1995).

\bibitem{Vobornik1999}
I. Vobornik et al., Physica C {\bf 317}, 589 (1999).

\bibitem{Chakravarty1993}
S. Chakravarty, A. Sudb\o, P. W. Anderson, and S. P. Strong, Science {\bf 261}, 337 (1993).

\bibitem{Chen2012}
C. Chen, A. Fujimori, C. Ting, and Y. Chen, arxiv. 1211.3477 (2012).

\bibitem{Bunemann2012} 
J. B\"unemann, T. Schickling, and F. Gebhard, Europhys. Lett. {\bf 98}, 27006 (2012).

\bibitem{Gebhard1990}
F. Gebhard, Phys. Rev. B {\bf 41}, 9452 (1990).

\bibitem{Kaczmarczyk2013}
J. Kaczmarczyk, J. Spa\l ek, T. Schickling, and J. B\"unemann, Phys. Rev. B {\bf 88}, 115127 (2013).

\bibitem{Kaczmarczyk2014}
J. Kaczmarczyk, J. B\"unemann, and J. Spa\l ek, New J. Phys. {\bf 16}, 073018 (2014).

\bibitem{Spalek2016}
J. Spa{\l}ek, M. Zegrodnik, and J. Kaczmarczyk, Phys. Rev. B {\bf 95}, 024506 (2017).

\bibitem{Ubbens1994}
M. Ubbens, P. A. Lee, Phys. Rev. B {\bf 50}, 438 (1994).

\bibitem{Lee1995}
P. A. Lee, K. Kuboki, J. Phys. Chem. Solids {\bf 56}, 1633 (1995).

\bibitem{Andersen1994}
O. K. Andersen, O. Jepsen, A. I. Lichtenstein, and I. I. Mazin, Phys. Rev. B {\bf 49}, 4145 (1994).

\bibitem{Jedrak2011}
J. J\k{e}drak and J. Spa\l ek, Phys. Rev. B {\bf 83}, 104512 (2011).

\bibitem{Kaczmarczyk2011}
J. Kaczmarczyk and J. Spa\l ek, Phys. Rev. B {\bf 84}, 125140 (2011).

\bibitem{Zegrodnik2014}
M. Zegrodnik, J. B\"unemann, and J. Spa\l ek, New J. Phys. {\bf 16},  033001 (2014).

\bibitem{Abram2016}
M. Abram, M. Zegrodnik, and J. Spa\l ek, Arxiv: 1607.05399 (2016).

\bibitem{Chen2012_2}
W.-Q. Chen, J. Y. Gan, T. M. Rice, and F. C. Zhang, Europhys. Lett. {\bf 98}, 57005 (2012).

\bibitem{Anderson1994}
P. W. Anderson, Physics B {\bf 199\& 200}, 8 (1994).

\bibitem{Hettinger1995}
J. D. Hettinger et al., Phys. Rev. Lett. {\bf 74}, 4726 (1995).

\bibitem{Wysokinski2015}
M. M. Wysoki\'nski, J. Kaczmarczyk, and J. Spa\l ek, Phys. Rev. B 92, 125135 (2015).

\bibitem{Wysokinski2016}
M. M. Wysoki\'nski, J. Kaczmarczyk, and J. Spa\l ek, Phys. Rev. B {\bf 94}, 024517 (2016).

\bibitem{Wang2006}
Q.H. Wang, Z.D. Wang, Y. Chen, and F.C. Zhang, Phys. Rev. B {\bf 73}, 092507 (2006).

\bibitem{Kaczmarczyk2014phmag}
J. Kaczmarczyk, Phil. Mag. 95, 563 (2014).

\bibitem{Byczuk1995}
K. Byczuk and J. Spa\l ek, Phys. Rev. B {\bf 53}, R518 (1996).

\bibitem{Furnier2010}
D. Furnier et al., Nature Phys. {\bf 6}, 905 (2010).

\bibitem{Yuli2008}
O. Yuli, I. Asulin, O. Millo, D. Orgad, L. Iomin, and G. Koren, Phys. Rev. Lett. {\bf 101}, 057005 (2008).

\bibitem{Anderson2007}
P. W. Anderson, Science {\bf 317}, 1705 (2007).

\bibitem{Coldea2001}
R. Coldea, S. M. Hayden, G. Aeppli, T. G. Perring, C. D. Frost, T. E. Mason, S.-W. Cheong, and Z. Fisk, Phys. Rev. Lett. {\bf 2001}, 5377 (2001).

\bibitem{Chakravarty1989}
S. Chakravarty, B. I. Halperin, and D. R. Nelson, Phys. Rev. B {\bf 39}, 2344 (1989).

\bibitem{Zhou2003}
X. J. Zhou \textit{et al.}, Nature {\bf 423}, 398 (2003).

\bibitem{Kordyuk2005}
A. A. Kordyuk, S. V. Borisenko, A. Koitzsch, J. Fink, M. Knupfer, and Berger, Phys. Rev. B {\bf 71}, 214513 (2005).

\bibitem{Nishida2010}
Y. Nishida, Phys. Rev. A {\bf 82}, 011605(R) (2010).

\bibitem{Klawunn2010}
M. Klawunn, A. Pikovski, L. Santos, Phys. Rev. A {\bf 82}, 044701 (2010).


\end{thebibliography}
\end{document}